\documentclass[preprint,showpacs,preprintnumbers,amsmath,amssymb]{revtex4}

\makeatletter
\def\@dotsep{4.5}
\makeatother
\usepackage{graphicx}
\usepackage{dcolumn}
\usepackage{bm}

\begin{document}



\title{Uncovering the  dynamics of citations of scientific papers}
\author{Michael Golosovsky\footnote{electronic address: michael.golosovsky@mail.huji.ac.il} and Sorin Solomon}
\affiliation{The Racah Institute of Physics, The Hebrew University of Jerusalem, 91904 Jerusalem, Israel\\}
\date{\today}

\begin{abstract}
We demonstrate a comprehensive framework that accounts for citation dynamics of scientific papers and  for the age distribution of references.  We show that citation dynamics of scientific papers is nonlinear and this nonlinearity has far-reaching consequences, such as diverging citation distributions and runaway papers. We propose a nonlinear stochastic dynamic model of citation dynamics based on link copying/redirection mechanism. The model is fully calibrated by empirical data and does not contain free parameters. This model can be a basis for quantitative probabilistic prediction of citation dynamics of individual papers and of the journal impact factor.

\end{abstract}
\maketitle

\section{Introduction}

The growth mechanism of complex networks is frequently attributed to preferential attachment \cite{Barabasi}. While  this mechanism accounts for the ubiquity of networks that are scale-free or have heavy-tailed degree distribution, it is too general and  does not specifically address evolving network structure. A more realistic scenario of the dynamics of growing networks is provided by the two-step growth models that have been developed in the context of social networks \cite{Jackson,Pennock}, epidemic-like propagation of ideas \cite{Goffman,Scharnhorst,Bettencourt},  diffusion of innovations \cite{Bass,Lal}, and citation dynamics \cite{Vitanov}.  In the context of citations these models are known  as redirection/copying \cite{Redner2005}, recursive search \cite{Vasquez2001,Vasquez2003}, link copying/referral \cite{Simkin},  uniform/preferential attachment \cite{Peterson}, and triad formation \cite{Holme2009,Ren}. Although citation network is specific (it is ordered, directed, acyclic, and does not allow rewiring \cite{Bagrow,Newman2009}), it is  an excellent example of a growing network since  it is well-documented and its dynamics can be reliably traced  through long time periods.

We introduce a comprehensive two-step model of a growing citation network. The model is fully calibrated by empirical data and does not contain free parameters. Our  measurements revealed an unexpected dynamic nonlinearity that was missing in all previous models. We incorporate this nonlinearity into our framework and come out with a nonlinear stochastic model of citation dynamics. The model predictions are confirmed by the measurements  of the age composition of the average reference list on the one hand, and  by the statistical distribution of cumulative citations for a large ensemble of papers, on another hand.

Our model can be useful for making probabilistic prediction of citations of scientific papers.  This active topic was initiated  by Refs. \cite{Penner,Acuna,Mazloumian,Barabasi2013,Uzzi} who suggested  several linear predictive  models  containing empirical parameters. Statistical uncertainty of these one-step models is too high. We introduce here a much more realistic nonlinear two-step model where all parameters have been calibrated in the independent measurements. The nonlinearity leads to divergent citation dynamics that explains why predicting citation behavior of individual papers is so difficult. Our model can be a basis for the probabilistic forecasting the scientific impact of a paper or of a journal.

\section{Statistics of references}
\subsection{Scenario: how an author composes his reference list}
\begin{figure*}[ht]
\begin{center}
\includegraphics[width=0.4\textwidth]{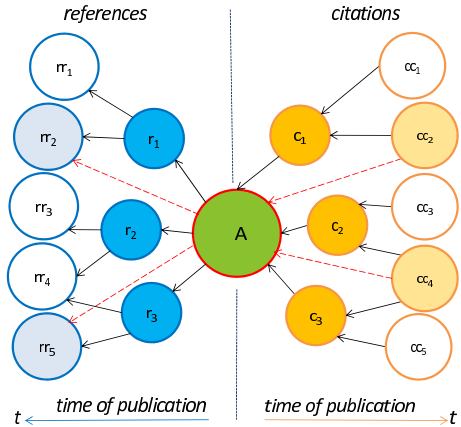}
\caption{ A fragment of a citation network showing two generations of references and citations of some parent paper A. The circles depict the papers, the solid lines depict direct references (citations), the dashed lines depict indirect references (citations). Each indirect reference (citation) closes a triangle  where the parent paper A is a vertex. The papers $r_{1}, r_{2},r_{3}$  are direct references of the paper A. The second-generation papers $rr_{1},rr_{2},..rr_{5}$  appear in the reference lists of the first-generation papers $r_{1},r_{2},r_{3}$. Some of the former ($rr_{2},rr_{5}$) also appear in the reference list of the parent paper A and we call them indirect references.    The papers $c_{1},c_{2},c_{3}$ (first generation citing papers) cite paper A directly. The papers $cc_{1},cc_{2},..cc_{5}$ (second generation citing papers) cite the first generation citing papers $c_{1},c_{2},c_{3}$. Some of the former ($cc_{2},cc_{4}$) also cite the parent paper A  and we call them indirect citations.
}
\label{fig:network}
\end{center}
\end{figure*}

Consider a cartoon scenario of an author writing a scientific paper. He reads research journals or media articles, searches the databases, finds the relevant papers and cites some of them in his reference list.  Then he studies the reference lists of these preselected papers, picks up  relevant references, reads them, cites some of them, and the process continues recursively.  We distinguish between the direct references that the author found through media or database search, and indirect references that the author picked up from the reference lists of the preselected papers  \cite{Vitanov,Simkin,Peterson}. Figure \ref{fig:network} shows the  corresponding reference network.

The direct and indirect references emerge in another scenario where the author finds each reference independently. Since old references are usually seminal studies, the author's most recent references will probably cite these old  papers as well. In our parlance the older references are indirect ones although the author could choose them without knowing that other preselected papers  cite them as well.

\subsection{Age distribution of references}
\begin{figure*}[ht]
\begin{center}
\includegraphics[width=0.4\textwidth]{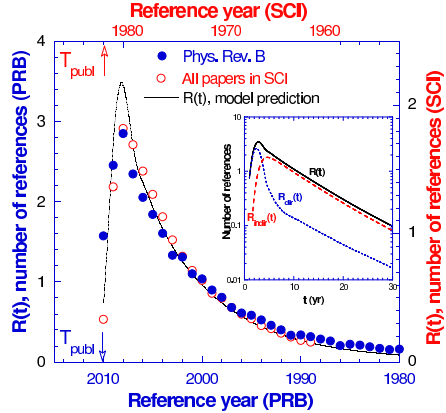}
\caption{$R(t)$,  a number of references in the reference list of a typical paper, that were published in a certain year $t$.  Blue circles show an average over all  papers published in Physical Review B in 2010 (excluding overviews). Red circles  show corresponding data for all papers covered by Science Citation Index that were published in 1982 (Ref.\cite{Nakamoto}). The arrows show publication year of the parent papers (2010 for the PRB papers and 1982 for the SCI papers).  Both dependences are almost identical (the difference in $y-$scales is due to the fact that the average length of the reference list in 2010 was $R_{0}=35$ while  $R_{0}=21.5$ in 1982). This identity shows that the age composition of the average reference list does not depend on the publication year of the parent paper. The solid blue line shows model prediction based on Eq. \ref{ref} with $R_{0}=30$ and $T=6.6e^{-0.64(t-1)}$ where $t=1$ is the publication year. The  inset shows model prediction for $R(t), R_{dir}(t), R_{indir}(t)$- the total, direct, and indirect references, correspondingly (see Eq. \ref{ref}).
}
\label{fig:ref}
\end{center}
\end{figure*}

The above scenario yields a very specific age distribution of references. Indeed, consider a reference list of an average paper that comprises $R_{0}=\int_{0}^{\infty}R(t_{0}, t_{0}-t)dt$ references where $t_{0}$  is the publication year  and  $R(t_{0}, t_{0}-t)$ is the number of references that were published in the year $t_{0}-t$. The latter  consist of the direct and indirect references,  $R(t_{0}, t_{0}-t)=R_{dir}(t_{0},t_{0}-t)+R_{indir}(t_{0},t_{0}-t)$.  For example, Fig. \ref{fig:ref} shows $R(t), R_{dir}(t), R_{indir}(t)$ for Physics papers published in $t_{0}=1984$.

To find $R(t)$ we make a crude approximation   that once the author cites some paper, he can  cite  any of its references \emph{with equal probability}.  An average reference list comprises $R(t_{0},t_{0}-\tau)$  preselected papers published in year $t_{0}-\tau$ (where $\tau <t$), each of which bringing in average $T(\tau)$ indirect references. The fraction of the latter  that were published  in the year $t_{0}-t$ is  $\frac{R(t_{0}-\tau,t_{0}-t)}{R_{0}(t_{0}-\tau)}$. This is equal to $\frac{R(t_{0},t_{0}-t+\tau)}{R_{0}(t_{0})}$ since the age composition of the average reference list is fairly independent of the publication year (Fig. \ref{fig:ref}). Finally, we obtain
\begin{equation}
\begin{split}
R(t)=&R_{dir}(t)+\int_{0}^{t}R(\tau)\frac{T(\tau)}{R_{0}}R(t-\tau)d\tau.\\
&\text{\small{direct\;\;\;\;\;\;indirect}}
\end{split}
\label{ref}
\end{equation}
Since all variables now  refer to the same  publication year $t_{0}$, we can drop $t_{0}$ from our notation, in such a way that $R(t)$  in  Eq. \ref{ref} denotes  $R(t_{0},t_{0}-t)$.

Once we know the functions $R_{dir}(t)$ and $T(t)$, we can solve Eq. \ref{ref} and find $R(t)$.    These functions  are the key parameters of the model since they capture  the citation habits of an average author. Although these functions could be found by analyzing reference lists of papers, this is not easy since bibliometric databases focus more on citations than on references. However, since there is a duality between citations and references, we can find $R_{dir}(t)$ and $T(t)$  by considering citation dynamics of the papers.

\section{Citation dynamics- a mean-field model}
\subsection{Reference-citation duality}
\begin{figure*}[ht]
\begin{center}
\includegraphics[width=0.35\textwidth]{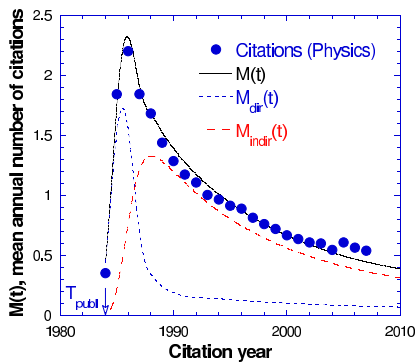}
\caption{$M(t)$, the mean annual number of citations garnered by a typical Physics paper published in 1984 (blue circles). The black line shows model prediction based on Eq. \ref{mean} with $R_{0}=21.5$ and $T=6.6e^{-0.64(t-1)}$ where $t=1$ is the publication year. The dashed lines show  the direct $M_{dir}(t)$ and indirect $M_{indir}(t)$ citations. $M_{dir}(t)$ shoots up 1-2 years after publication and then  sharply decays, while $M_{indir}(t)$ achieves its maximum after 3-4 years and then slowly decays.
}
\label{fig:mean}
\end{center}
\end{figure*}

Since one paper's citation is another paper's reference,  the reference and citation networks are dual (Fig. \ref{fig:network}). Consequently, the age distribution of references $R(t)$ for the papers published in one year (diachronous  or retrospective citation distribution \cite{Nakamoto,Glanzel}) (Fig. \ref{fig:ref}) is very similar to  $M(t)$ (Fig. \ref{fig:mean}), the mean citation rate of the papers published in one year (synchronous or prospective citation distribution).

In what follows we analyze consequences of this duality and how it can be used to measure relevant parameters in Eq. \ref{ref}.  Indeed,  consider a set of all $N_{0}(t_{0})$  papers in a certain research field that were published in  year $t_{0}$.  The mean number of citations that a paper from this set garners in the $t$-th year after publication is $M(t_{0},t_{0}+t)$. Since the majority of citing papers belong to the same research field, the total number of citations garnered by these $N_{0}(t_{0})$ papers in the year $t_{0}+t$ shall be equal to the total number of the  references in the reference lists of the papers published in the year $t_{0}+t$,
\begin{equation}
N_{0}(t_{0})M(t_{0},t_{0}+t)= N_{0}(t_{0}+t)R(t_{0}+t,t_{0})
\label{duality}
\end{equation}
Equation \ref{duality} relates the total number of citations and references for the papers  belonging to the same research field but published  in different years. To find corresponding relation for the papers published in the same year, we shall take into account  that both the number of publications $N_{0}$ and the reference list length $R_{0}$  grow  exponentially with time, $N_{0}(t_{0})\propto e^{\alpha t_{0}}, R_{0}(t_{0})\propto e^{\beta t_{0}}$. We substitute these exponential  dependences into Eq. \ref{duality},  notice that $\frac{R(t_{0}+t,t_{0})}{R_{0}(t_{0}+t)}=\frac{R(t_{0},t_{0}-t)}{R_{0}(t_{0})}$, and find the mathematical expression for the reference-citation duality,
\begin{equation}
M(t_{0},t_{0}+t)=e^{(\alpha+\beta)t}R(t_{0},t_{0}-t).
\label{duality1}
\end{equation}
\subsection{ A mean-field model of citation dynamics}
The substitution of  Eq. \ref{duality1}  into Eq. \ref{ref} yields dynamic equation for the mean citation rate of the papers published in one year
\begin{equation}
\begin{split}
M(t)=&M_{dir}(t)+\int_{0}^{t}M(t-\tau)\frac{T(t-\tau)}{R_{0}} M(\tau)d\tau\\
&\text{\small{direct\;\;\;\;\;\;indirect}}
\end{split}
\label{mean}
\end{equation}
where  $M_{dir}=R_{dir}(t)e^{(\alpha+\beta)t}$ and $T(\tau)$ has been replaced by $T(t-\tau)$ using the properties of the convolution.  Equation \ref{mean} tells us that an average  paper published in some year $t_{0}$  has  $M(\tau)$  first-generation citations published in year $t_{0}+\tau$,  each of which  generating $M(t-\tau)$ second-generation citations in some later year $t_{0}+t$.  The probability that a second-generation citation induces (indirect) citation of the parent paper is $T(t-\tau)/R_{0}$ where $R_{0}$ is the average length of the reference list of the papers published in the year $t_{0}$.

To solve Eq. \ref{mean} we have to determine functions $M_{dir}(t)$ and $T(t)$. In fact, we used Eq. \ref{mean} to find $T(t)$. To this end we measured $M(t)$ and  $M_{dir}(t)$ (see next section), substituted these functions into Eq. \ref{mean}, and found an exponential kernel $T=T_{0}e^{-\gamma (t-1)}$ where $T_{0}=6.6$, $\gamma=0.64$ yr$^{-1}$, and the publication year corresponds to $t=1$. (These numbers mean that  $\sim 35\%$ references in the average reference list of a paper are direct and $\sim 65\%$ are indirect).

To find age distribution of references $R(t)$ we  calculated $R_{dir}(t)=M_{dir}(t)e^{-(\alpha+\beta)t}$ where for Physics papers $\alpha=0.046,\beta=0.02$ (as found in our independent measurements). We substituted $R_{0}=35$, and the functions $R_{dir}(t), T(t)$ found above  into Eq. \ref{ref} and solved it to find $R(t)$. Figure \ref{fig:ref} shows that the model prediction $R(t)$  fits  perfectly well our measurements.   We conclude that the mean-field model (Eqs.\ref{ref},\ref{duality},\ref{mean}) faithfully accounts for the average age composition of the reference list and for the mean citation  dynamics of scientific papers.  In what follows we use this model to infer citation dynamics of individual papers.

\section{Citation dynamics of individual papers}
\subsection{Linear stochastic model}
To infer equation describing citation dynamics of individual papers we consider $\Delta k^{A}$, the number of citations garnered by some paper $A$ during a short time interval from $t$ to $t+\Delta t$. The cumulative number of citations of this paper is $k^ {A}(t)=\sum_{0}^{t}{\Delta k^{A}}$.  We assume that $\Delta k^{A}$ is a discrete random variable that follows a time-inhomogeneous Poisson process  \cite{Golosovsky}  with the rate $\lambda^{A}(t)$. This latent citation rate \cite{Burrell} consists of the direct and indirect contributions, $\lambda^{A}(t)=\lambda^{A}_{dir}(t)+\lambda^{A}_{indir}$.

To infer  dynamic equation for $\lambda^{A}$ we note that for the  set of papers published in one year, the average rate of direct citations is  $M_{dir}(t)=\overline{\lambda^{A}_{dir}(t)}$, the average rate of total citations is $M(t)=\overline{\lambda^{A}(t)}$, and
$M(\tau)d\tau=\overline{\Delta k^{A}(\tau)}$ where $d\tau=1$. We substitute these equalities into Eq. \ref{mean}, replace integral by sum, dispense with the averaging, and obtain
\begin{equation}
\begin{split}
\lambda^{A}(t)=&\lambda^{A}_{dir}(t)+\sum_{\tau=0}^{t}M(t-\tau)\frac{T(t-\tau)}{R_{0}}\Delta k^{A}\\
&\text{\small{direct\;\;\;\;\;\;indirect}}
\end{split}
\label{paper}
\end{equation}
This discrete stochastic equation is consistent with Eq. \ref{mean}. Our initial assumption (to be revised soon) is that the functions $M(t-\tau)$ and $T(t-\tau)$, determined from our  studies of mean-field citation dynamics, govern citation dynamics of individual papers as well.

\subsection{Measurements and comparison to the model}
To verify Eq. \ref{paper} empirically we need to measure the direct and indirect citations separately. To this end we chose 37 representative research papers that were published in the  Physical Review B  in one  year,  analyzed their first- and second-generation citing papers (Fig. \ref{fig:network}), identified  the direct and indirect citations and measured their dynamics. As an aggregate measure of the paper's  individuality we took $k_{\infty}$, the long-time limit of cumulative citations.
\subsubsection{Direct citations}
We found (not shown here) that the direct citation rate of a paper can be represented as
\begin{equation}
\lambda^{A}_{dir}(t)=p^{A}m(t)
\label{direct}
\end{equation}
where $p^{A}$ is the numerical parameter and the function $m(t)$ is the same  for all papers published in one year, whereas  $\int_{0}^{\infty}m(t)dt=1$. Figure \ref{fig:mean} shows that $m(t)$ grows immediately after publication of the paper, achieves its maximum after $\sim$ 2 years and slowly decays thereafter. The long tail of  $m(t)$ is a mathematical expression of  delayed recognition  ("sleeping beauty" \cite{sleeping-beauty}) phenomenon.

The parameter $p^{A}$ is the long-time limit of the  number of direct citations and it is a proxy to the so-called "fitness"  \cite{Simkin,Bianconi} which shall depend on the scientific quality of the parent paper, the journal where it was published,  popularity of the research field, etc.  On the one hand, $p^{A}$  can be  estimated  \emph{a priori} from the initial citation rate of the paper. [Indeed,   shortly after publication $\frac{dk^{A}}{dt}|_{t=1}\approx \lambda_{dir}|_{t=1}=p^{A}m|_{t=1}$.]
On another hand, since the solution of Eqs. \ref{paper},\ref{direct} yields $k^{A}_{\infty}\propto p^{A}$, this relation can serve as an \emph{a posteriori} estimate of $p^{A}$. Our  measurements (not shown here) yield a sublinear dependence
\begin{equation}
p^{A}=0.72(k^{A}_{\infty}+k_{0})^{0.8}
\label{fitness}
\end{equation}
where small parameter $k_{0}\approx 1$  accounts for the fact that previously  uncited papers have some probability to be cited in future.

\subsubsection{Indirect citations}
We found that Eq. \ref{paper} with the kernel  $\frac{T_{0}}{R_{0}}M(t-\tau)e^{-\gamma(t-\tau)}$   fits the dynamics of indirect citations of individual papers only if we allow for $T_{0}$ and  $M$  to depend on the number of previous citations $k$. The reason  for this surprising $k$-dependence is that new  citations modify the very structure of  citation network associated with the cited paper, this modification being most pronounced for highly-cited papers.

\begin{figure*}[ht]
\includegraphics[width=0.45\textwidth]{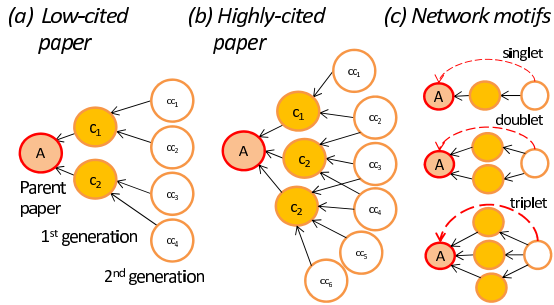}
\begin{center}
\caption{(a)  Two generations of the papers citing a low-cited parent paper. Each second-generation paper cites only one first-generation paper and it is connected to the parent paper by a single path. The numbers of the second-generation citations $M$ and citing papers $N$ are equal, $r=M/N=1$. (b)  Two generations of the papers citing a highly-cited parent paper. Each second-generation paper can cite several first-generation papers and it can be connected to the parent paper by multiple paths ($cc_{1},cc_{5},cc_{6}$ are connected to the parent paper by a single path; $cc_{2},cc_{3},cc_{4}$ are connected to the parent paper by  double paths). The numbers  of second-generation citations  and  citing papers are not equal, $r=M/N=1.5$. (c) Network motifs. Solid lines show direct citations, dashed line show indirect citations. The probability of indirect citation progressively  increases from singlet to doublet to triplet as $\approx 1:4:9$ indicating constructive multipath interference.
}
\label{fig:generations}
\end{center}
\end{figure*}

\begin{figure*}[ht]
\includegraphics[width=0.4\textwidth]{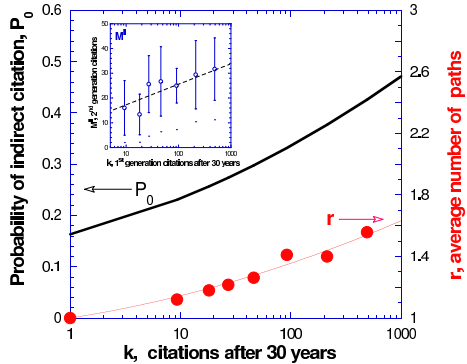}
\begin{center}
\caption{Microscopic parameters of citation dynamics and their dependence on the number of cumulative  citations $k$.   The solid circles show $r=M/N$, the ratio of the numbers of the second- generation citations $M$ to second-generation citing papers $N$ (analysis of 108 PRB papers published in 1984). $r$ characterizes the average number of paths connecting a second-generation citing paper to the parent paper. $r$ increases with $k$ following empirical dependence $r=1+0.11\log k+0.033(\log k)^2$ (red solid line). The inset shows that $M^{II}$, the number of the second-generation citations per one first-generation citing paper, slowly increases with $k$.   The solid black line shows $P_{0}$, the probability of indirect citation of the parent paper by a second-generation citing paper. It follows functional dependence $P_{0}=a[1+b(r-1)]$ derived from our model where parameters $a=0.16$ and $b=3$ were found from the  microscopic measurements.
}
\label{fig:ratio}
\end{center}
\end{figure*}

Indeed, consider two generations of citing papers associated with a parent paper. Obviously, the number of the first-generation citations $k$ is equal to the number of the first-generation citing papers. However, the numbers of second generation citations  and citing papers can differ. We denote by $M$ and $N$, correspondingly, the long-time limits of the number of second-generation citations and citing papers per one first-generation citing paper, and introduce $r=M/N$, an average number of the first-generation citing papers cited by a second-generation citing paper.  Figure \ref{fig:ratio} shows that  $r$ increases with $k$  following empirical dependence $r=1+0.11\log k+0.033(\log k)^2$.  The inset shows that this growth is associated with the $M(k)$ dependence (this means that the  citation network is assortative) while $N$ is almost independent on $k$. Figure \ref{fig:generations} demonstrates that $r$ measures  the average number of paths leading from the parent paper to a second-generation citing paper. For a low-cited parent paper $r=1$, indicating that it is connected to each of its second-generation descendant by a single path. For a highly-cited parent paper  $r>1$ indicating that it is connected to some of its second-generation descendants by multiple paths.

We found that the parameter $T_{0}$ in Eq. \ref{paper} is also $k$-dependent (not shown here). Therefore,  we merge all $k$-dependent parameters together and introduce $P_{0}=rT_{0}/R_{0}$, the probability  that a second-generation citing paper cites the parent paper (indirectly). Our measurements revealed that $P_{0}$ increases nonlinearly with $j$, the number of citation paths connecting the parent paper with its second-generation descendant, as it is schematically shown in Fig. \ref{fig:generations}c. In particular, we found quadratic dependence $P_{0}\propto j^{2}$   indicating constructive interference between multiple paths. Since the number of multiple paths increases with $k$, this translates into empirical dependence $P_{0}(k)= 0.16[1+3(r-1)]$ (while in the absence of multipath interference one would obtain $P_{0}(k)= 0.16r$)  The $P_{0}(k)$ dependence stems from the $r(k)$ dependence and it has loose analogy with  bootstrap percolation.


\section{Nonlinear model of citation dynamics}
\subsection{Dynamic equation}
To introduce nonlinearity  into Eq. \ref{paper} we replaced the kernel $\frac{T_{0}}{R_{0}}M(t-\tau)e^{-\gamma(t-\tau)}$ by $P_{0}(k)N(t-\tau)e^{-\gamma(t-\tau)}$. Here, $N(t)$ is the average  number of the second generation citing papers per one first-generation citing paper (fan-out coefficient), and $P_{0}$  is the probability that a second-generation citing paper cites the parent paper (indirectly). The novelty here is the $P_{0}(k)$ dependence which is shown in  Fig. \ref{fig:ratio}. We introduce this kernel  into Eq. \ref{paper}, plug there Eq. \ref{direct}, and obtain our key result- nonlinear stochastic dynamic equation for the latent citation rate of a paper A-
 \begin{equation}
\lambda^{A}(t)=p^{A}m(t)+\sum_{\tau=0}^{t}P_{0}(k^{A})e^{-\gamma(t-\tau)}N(t-\tau)\Delta k^{A}(\tau)
\label{paper-new}
\end{equation}
The empirical functions $m(t)$, $N(t)$, and $P_{0}(k)$ are shown in the Figs. \ref{fig:mean},\ref{fig:ratio}, correspondingly. Equation \ref{paper-new} is a nonlinear first-order discrete stochastic differential equation with the initial condition set by $p^{A}$.  This equation expresses $\lambda^{A}(t)$, the latent citation rate of the paper $A$ at time $t$, through  past citations of the same paper, $\Delta k^{A}(\tau)$ and $k^{A}=\sum{\Delta k^{A}(\tau)}$. The probability distribution of actual citations at time $t$ is given by the Poisson distribution, $P(\Delta k)=\frac{(\lambda^{A})^{\Delta k}}{(\Delta k)!}e^{-\lambda^{A}}$. 


\subsection{Stochastic simulation}
\begin{figure*}[ht]
\begin{center}
\includegraphics[width=0.45\textwidth]{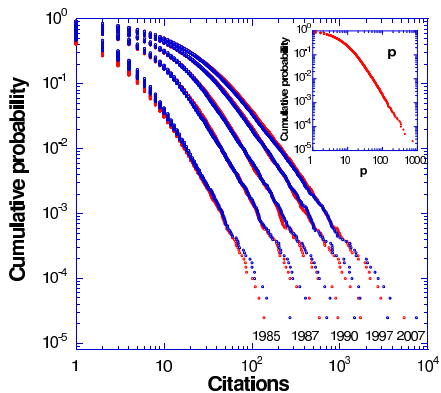}
\caption{ Cumulative citation distributions for  40,195 Physics papers published in 1984. Red symbols stay for measurements, blue symbols stay for the  results of a stochastic simulation based on  Poisson process with the rate given by Eq. \ref{paper-new} and the probability $P_{0}$ shown in Fig. \ref{fig:ratio}. The inset shows fitness $p^{A}$ estimated from  Eq. \ref{fitness}.
}
\label{fig:simulation}
\end{center}
\end{figure*}

To verify Eq. \ref{paper-new} we performed  stochastic  numerical simulation  imitating  citation dynamics of a set of 40195 Physics papers published in  1984.  Figure \ref{fig:simulation} shows the cumulative citation distributions for this set over the time span of 25 years. We wish to imitate these distributions using Eq. \ref{paper-new}. This requires that the statistical distribution of initial conditions ("fitness"  $p^{A}$) for the actual and "simulated" papers be the same.   We estimated $p^{A}$ for each paper using Eq. \ref{fitness} and assuming $k_{\infty}\approx k(t=25)$. The inset in Fig. \ref{fig:simulation} shows corresponding statistical distribution of $p^{A}$.

We run stochastic simulation based on Eq. \ref{paper-new} with this distribution of $p^{A}$ and empirical functions $m(t)$, $N(t)$, $P_{0}(k)$  shown in Figs. \ref{fig:mean},\ref{fig:ratio}, correspondingly. Figure \ref{fig:simulation} shows excellent  agreement between the simulated and measured cumulative citation distributions.  Moreover, our simulation  accounts fairly well  for such intricate characteristics of citation dynamics as stochastic variability, temporal autocorrelation, and the dynamics of uncited papers (not shown here). We present here our measurements with Physics papers while we obtained very similar results with the Mathematics and Economics papers as well.

\subsection{Analysis of the model}
To have more insight into citation dynamics of scientific papers we consider continuous approximation of  Eq. \ref{paper-new}. Without  loss of generality  we  disregard stochasticity, consider $k$ as a continuous variable, and approximate the kernel by the exponential, $P_{0}(k)N(t-\tau)e^{-\gamma(t-\tau)}\approx q(k)e^{-\gamma'(t-\tau)}$ where  $q\approx 1.26P_{0}(k)$ and $\gamma'=\gamma+0.08$. [The rationale for this approximation is the fact that $e^{-\gamma(t-\tau)}$ with $\gamma=0.64$ yr$^{-1}$ has much stronger time dependence than $N(t-\tau)$. The latter is captured by the term 0.08 yr$^{-1}$]. We  replace the sum in Eq. \ref{paper-new}  by the integral, drop  index $A$ and arrive at
\begin{equation}
\frac{dk}{dt}=pm(t)+\int_{0}^{t}qe^{-\gamma'(t-\tau)}\frac{dk}{d\tau}d\tau
\label{dynamics2}
\end{equation}
Equation \ref{dynamics2} appears in the context of Bellman-Harris branching (cascade) processes \cite{Bellman}. It is  well-known in the population dynamics where it describes the age-dependent birth-death process with immigration \cite{Ebeling} where direct and indirect citations are analogs of immigration and reproduction, correspondingly, and $q/\gamma'$ is the reproduction number.

Dynamic behavior described by Eq. \ref{dynamics2} results from the interplay between the positive feedback  rate characterized by the factor $q$  and  the rate of obsolescence characterized by the parameter $\gamma'$. The latter shall be compared to the average paper longevity (citation lifetime), $\tau_{0}$, that we define empirically using a crude exponential approximation $k(t)=k_{\infty}[1-e^{-(t-\Delta)/\tau_{0}}]$, where $\Delta$ characterizes delayed recognition. In the limit  $\gamma'\tau_{0}>>1$ Eq. \ref{dynamics2} reduces to the first-order autoregressive model of citation dynamics \cite{Golosovsky}
\begin{equation}
\frac{dk}{dt}\approx pm(t)+\frac{q}{\gamma'}\left(\frac{dk}{dt} \right)_{t-1/\gamma'}
\label{dynamics-short}
\end{equation}

In the opposite limit, $\gamma'\tau_{0}<<1$, Eq. \ref{dynamics2}  reduces to the models of Refs. \cite{Redner2005,Peterson}
\begin{equation}
\frac{dk}{dt}\approx p m(t)+qk
\label{dynamics-Bass}
\end{equation}
The latter is nothing else but the Bass equation for diffusion of innovations \cite{Bass,Lal} in an infinite market. Citations correspond to adopters, direct citations correspond to innovators, and indirect citations correspond to imitators.  The connection to the Bass model is not occasional since each paper can be considered as a new product whose penetration to the market of ideas is gauged by the number of citations. The novelty here is the nonlinear $q(k)$ dependence. In the context of diffusion of innovations  the nonlinear coefficient of imitation $q(k)$  is not unexpected. This would indicate increased probability of adoption of a new product if several neighbors in the network already adopted it. To the best of our knowledge, such possibility didn't deserve much attention.

\section{Consequences of nonlinearity: runaways}
To analyze consequences of the nonlinearity we note that since $q(k)$ dependence is weak we can  integrate  Eq. \ref{dynamics2}  over time assuming constant $q$. This yields
\begin{equation}
k(t)\approx p\int_{0}^{t}\left[m(t')+q\int_{0}^{t'}m(\tau)e^{-(\gamma'-q)(t'-\tau)}d\tau \right]dt'
\label{dynamics-delta}
\end{equation}

The first term in the square brackets corresponds to direct citations, the second term stays for indirect citations. Each direct citation induces a cascade of indirect citations that propagates in time if $\gamma'-q<0$ and decays if  $\gamma'-q>0$. In the latter case $k$ comes to saturation, $k_{\infty}\rightarrow p \gamma'/(\gamma'-q)$ (ordinary papers) while in the former case $k$ grows exponentially  (seminal papers). Since $k$ grows with time, an ordinary paper which by pure chance garnered excessive number of citations, can become a seminal paper.

Although the $q(k)$ dependence is weak, it is important  since it enters in the exponent.  This results in a "winner takes all" instability \cite{Vasquez2003,Redner2001,Mondragon,Krap-kryukov,Barabasi2012}. To analyze how this instability develops with $k$ we again consider the paper longevity (citation lifetime) $\tau_{0}$. The latter is determined by the  exponent $\gamma'-q$, and to a lesser extent, by the function $m(t)$.  Equation \ref{dynamics-delta} suggests that $\tau_{0}\propto 1/(\gamma'-q)$. Since $q$ increases with $k$, the inverse  $\tau_{0}(q)$ dependence  means  that with increasing number of citations, $\tau_{0}$ increases and diverges  upon approaching the branching (tipping) point $q(k)=\gamma'$ \cite{Simkin}. This means that each new citation extends a paper's lifetime \footnote{Quite opposite to the situation described in the famous Balzac's novel "La peau de chagrin".}

\begin{figure*}[ht]
\begin{center}
\includegraphics[width=0.35\textwidth]{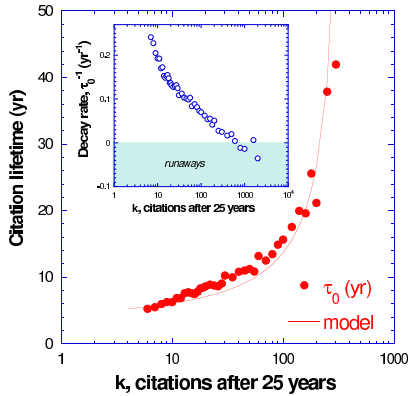}
\caption{The paper longevity (citation lifetime), $\tau_{0}$,  and the corresponding decay rate $\tau_{0}^{-1}$, versus $k$, the number of citations after 25 years. To reduce fluctuations the data were binned. $\tau_{0}$ increases with increasing $k$, in such a way that highly-cited papers show runaway behavior (diverging $\tau_{0}$, negative decay rate). The solid line shows a crude approximation, $\tau_{0}\propto \frac{1}{\gamma'-q(k)}$, suggested by Eq. \ref{dynamics-delta}. Here, $\gamma'=0.72$ yr$^{-1}$ and $q=1.26P_{0}(k)$ where $P_{0}(k)$ is shown in Fig. \ref{fig:ratio}. 
}
\label{fig:lifetime}
\end{center}
\end{figure*}

Figure \ref{fig:lifetime} demonstrates  that the citation lifetime $\tau_{0}$ indeed increases with increasing $k$ and diverges when $k>600$, in such a way that the papers with more than 600-1000 citations exhibit runaway behavior - their citation career does not saturate even after 25 years. This complements the famous parable "rich get richer" by  "rich live longer".

\section{Summary}
We developed a nonlinear stochastic model of citation dynamics of scientific papers and validated this model by measurements.  The underlying scenario is as follows. We assume that the author of a new scientific paper finds relevant papers from the media or journals and cites  them. Then he studies the reference lists of these preselected papers, picks up  some references, cites them as well, and continues this process recursively. We add here a new ingredient: if some paper is cited by several preselected papers,  the author chooses it with higher probability than that cited by only one preselected paper.

This new ingredient, combined with the assortativity of the citation network, introduces  dynamic nonlinearity. The  account of this nonlinearity is crucial for predicting future citation behavior of the papers. Our  nonlinear dynamic model can serve as a basis for probabilistic forecasting of citation dynamics of a paper or a group of papers (journal impact factor).

\begin{acknowledgments}
We are grateful to S. Redner, A. Scharnhorst, L. Muchnik, and D. Shapiro for fruitful discussions, we appreciate instructive correspondence with M. Simkin. We acknowledge financial support of the EU COST Action TD1210.
\end{acknowledgments}




\end{document}